\documentclass{elsart}

\usepackage{natbib}

 \usepackage{graphics}
\begin{document}

\begin{frontmatter}



\title{Highly extinguished supernovae in the nuclear regions of starburst galaxies}

\author[seppo]{S. Mattila}
\ead{seppo@astro.su.se}
\author[peter]{W.P.S. Meikle}
\author[robert]{R. Greimel}

\address[seppo]{Stockholm Observatory, AlbaNova, Stockholm SE-106 91, Sweden}
\address[peter]{Blackett Lab., Imperial College, Prince Consort Road, London SW7 2BW, UK}
\address[robert]{Isaac Newton Group of Telescopes, Apartado de correos 321, E-38700 Santa Cruz de la Palma,
Tenerife, Spain}

\begin{abstract}
A handful of nearby supernovae (SNe) with visual extinctions of a few magnitudes
have recently been discovered. However, an undiscovered population of
much more highly extinguished ($A_{V}$ $>$ 10) core-collapse
supernovae (CCSNe) is likely to exist in the nuclear (central kpc)
regions of starburst galaxies. The high dust extinction means that
optical searches for such SNe are unlikely to be successful. Here, we
present preliminary results from our ongoing near-infrared Ks-bands
search programme for nuclear SNe in nearby starburst galaxies.  We
also discuss searches for SNe in Luminous and Ultraluminous Infrared
Galaxies.
\end{abstract}


\end{frontmatter}

% main text
%%%%%%%%%%%%%%%%%%%%%%%%%%%%%%%%%%%%%%%%%
\section{Introduction: Obscured supernovae}
Core-collapse (types II and Ib/c, hereafter CCSNe) supernovae are observed
to occur in sites of recent star formation. Such regions contain large
quantities of dust, especially the nuclear (central kiloparsec) regions
of starburst galaxies. Consequently, SN search programmes working at
optical wavelengths most likely miss a significant number of events in
starburst galaxies. The distributions of the host galaxy extinctions for 
samples of nearby thermonuclear (type Ia) SNe, and CCSNe are shown in Fig.1. 
Although, most of the {\it discovered} SN events have extinctions below $A_{V}$ = 1, the
extinction distributions of both types of SNe show tails extending to
several magnitudes. 
\begin{figure}
\hspace{-0.7cm}
\resizebox{15cm}{!}{
{\includegraphics{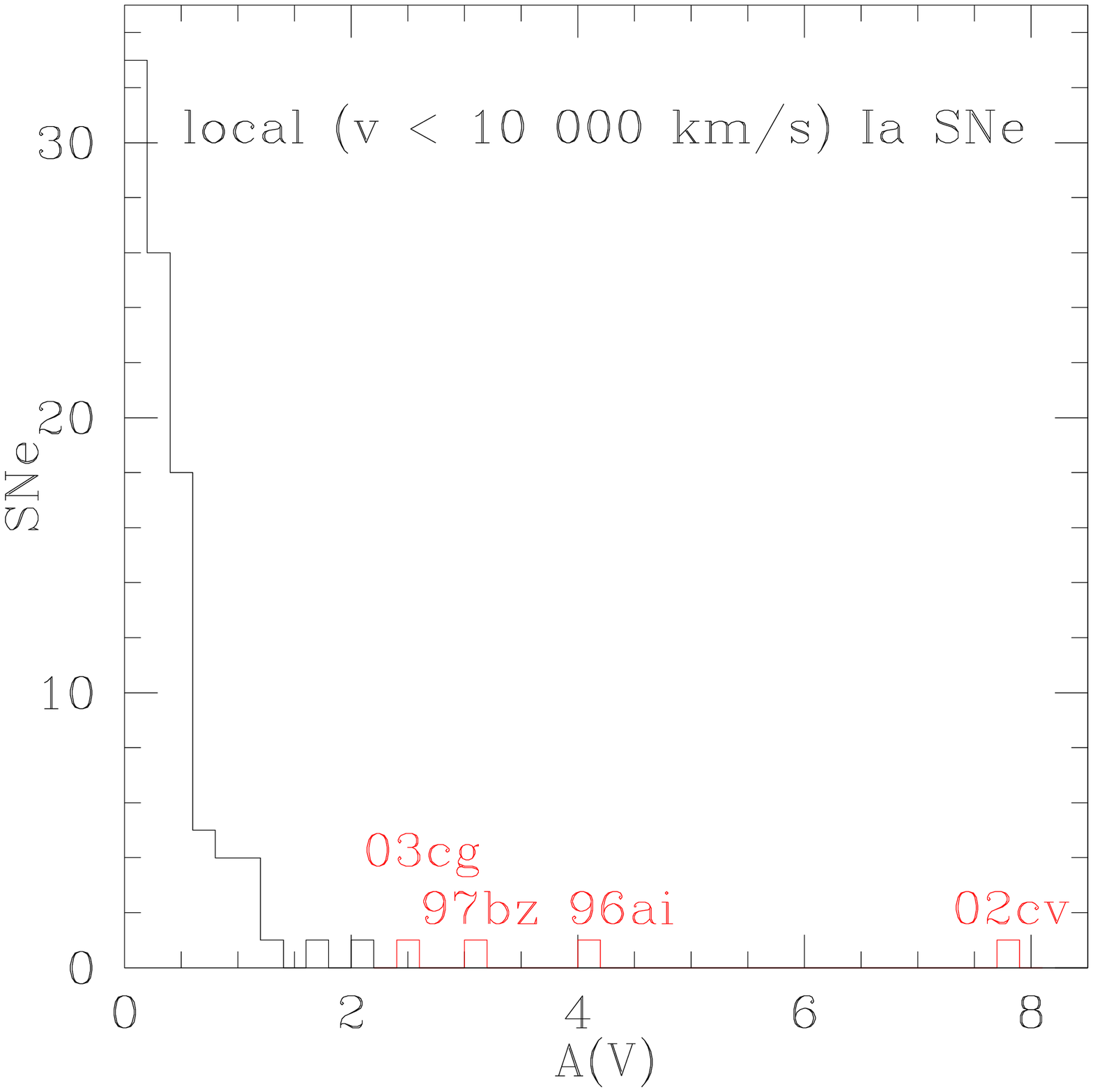}}
{\includegraphics{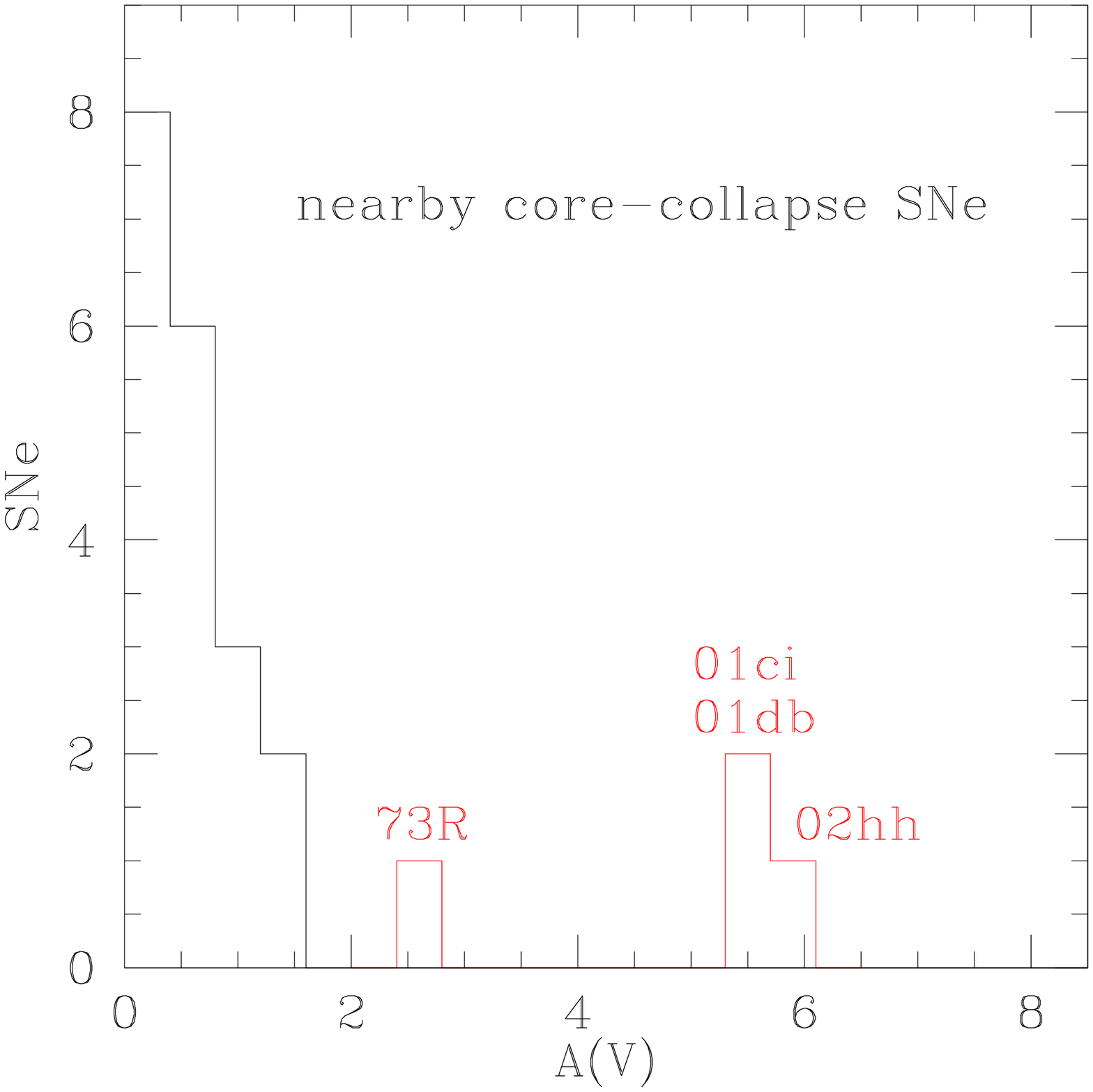}}
{\includegraphics{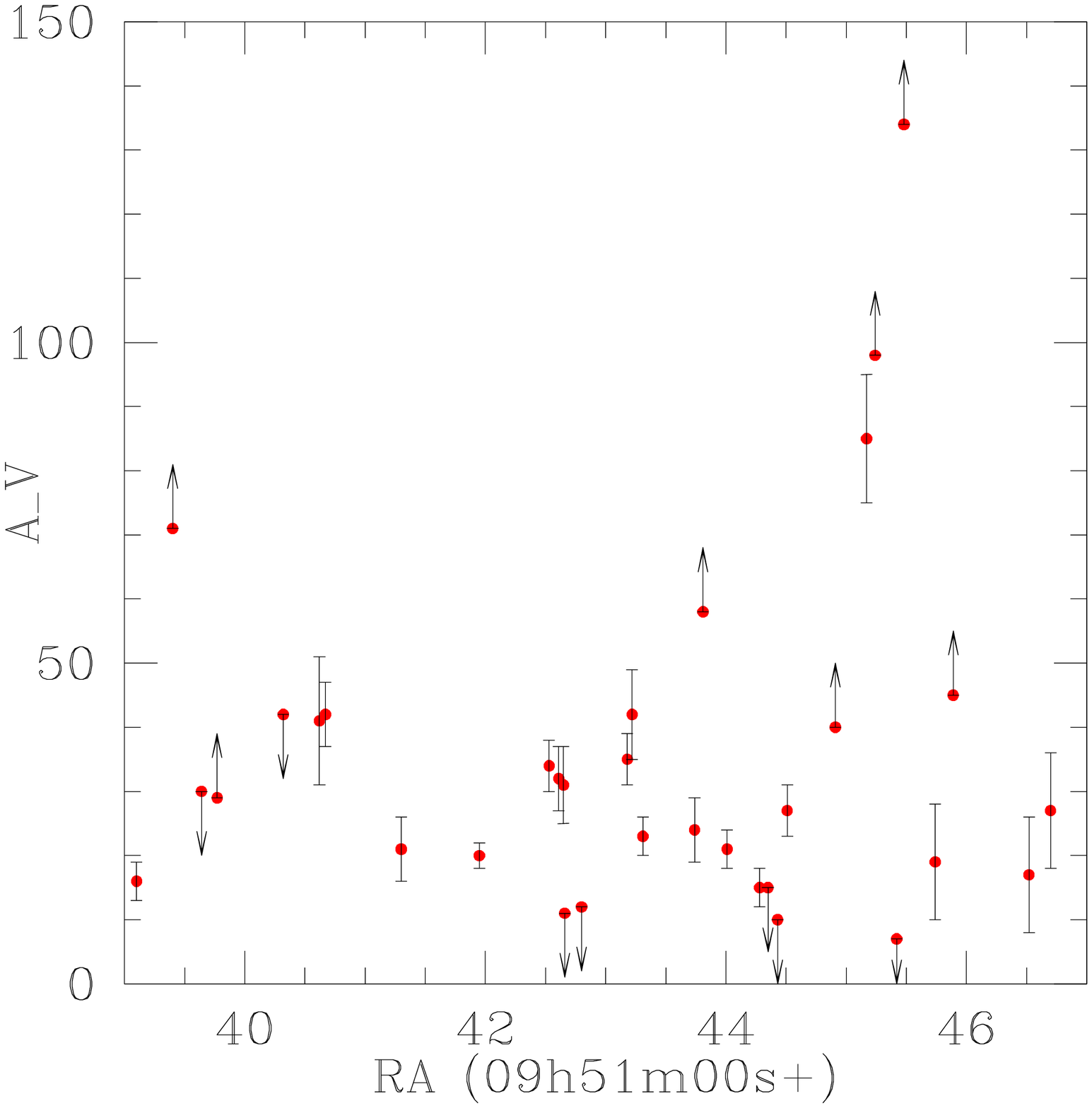}}
}
\caption{The host galaxy extinctions of nearby type Ia SNe (left)
[4,9,10], and CCSNe (middle) [1,2,5,11,12]. The extinctions towards the SNRs 
of M~82 (right) [12].}
\label{fig1}
\end{figure}
In particular, four nearby SNe with host galaxy
extinctions of several magnitudes ($A_{V}$ $>$ 5) [1,2,3,4,5] have been
discovered during the last couple of years.  The type II SNe 2001ci
[6] and 2002hh [7] were detected at optical wavelengths thanks to
their small distances of only 14 and 6 Mpc respectively. The more
distant type II SN 2001db [2] (37 Mpc) and type Ia SN 2002cv [8] (22
Mpc) were discovered in the near-infrared (IR). All these SNe have active
host galaxies (either H II or LINER/Seyfert 2) according to NED$\footnote{
The NASA/IPAC Extragalactic Database (NED) is operated by the Jet Propulsion 
Laboratory, California Institute of Technology, under contract with the National 
Aeronautics and Space Administration.}$. Three
of them have projected galactocentric distances smaller than $\sim$2
kpc, and one, SN 2002cv, is located behind an optical dust
lane. Recent optical spectra of SN 2002hh are shown in Fig.2.  The
effects of the high extinction ($A_{V}$ $\sim$ 6) [5] are clearly
visible as a dramatic drop of the signal towards the shorter
wavelengths (note also the lack of H$_{\beta}$ emission). However, an
as yet unrevealed population of much more highly extinguished CCSNe is
likely to exist in the nuclear (central kiloparsec) regions of
starburst galaxies. In the nuclear regions ($\sim$ 600pc diameter) of
the starburst galaxy M~82, large hydrogen column densities indicate
extinctions [12] of $A_{V}$ = 30 ($\sigma$ $\sim$ 16) (Fig.1 right)
towards the group of young supernova remnants (SNRs) observed at radio
wavelengths. Furthermore, recent VLBA monitoring [13,14] has revealed a
group of luminous radio SNe/SNRs within the nuclear regions ($\sim$100
and $\sim$200 pc diameter around the Eastern and Western nuclei
respectively) of the nearest Ultraluminous Infrared Galaxy (ULIRG)
Arp~220.  The estimates for the explosion rates of such {\it luminous
radio SNe} range between $\sim$0.5 and 2.0 yr$^{-1}$ [13,14]. However,
it is likely that not all the CCSNe in Arp~220 have suitable CSM/ISM
conditions to produce high radio luminosities i.e.  many SNe which
happen to explode in less extreme environments probably remain
undetected in the radio. Therefore, near-IR monitoring campaigns of
starburst galaxies having a range of IR luminosities should be carried
out to complement the rate estimates from the radio observations.
Near-IR SN searches were first attempted over a decade ago [15,16] but
it is only now, with the introduction of large format (1024 x 1024)
small pixel scale ($<$0.3'') near-IR detectors on 2-4 meter
telescopes, that such searches have become realistically feasible.
\begin{figure}
\begin{center}
\resizebox{12cm}{!}{
\rotatebox{-90}{
{\includegraphics{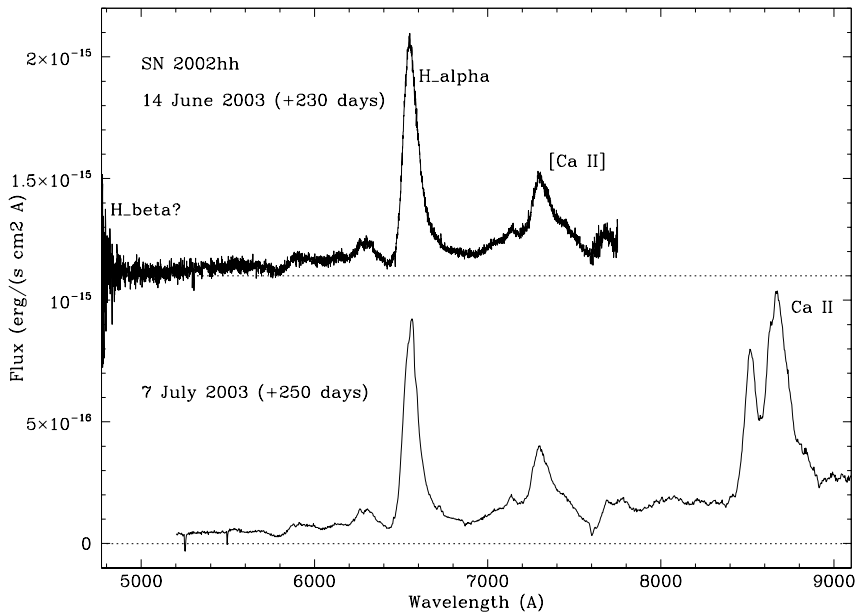}}} 
}
\caption{Optical spectra of the obscured ($A_{V}$ $\sim$ 6) type II SN 
2002hh observed with WHT/ISIS (14 June 2003) and NOT/ALFOSC (7 July 2003).  
The former spectrum has been shifted upward by 1.1 $\times$
10$^{-15}$ erg s$^{-1}$ cm$^{-2}$ \AA$^{-1}$ for clarity. The zero-levels
of the spectra are shown with dotted lines. The most prominent SN lines
are identified.
}
\label{fig1}
\end{center}
\end{figure}
\section{Supernovae in nearby starburst galaxies}
Since August 2001 we have been carrying out a near-IR search programme
for CCSNe in the nuclear regions of nearby starburst galaxies [17]. We use
the INGRID near-IR camera at the William Herschel Telescope (WHT) on
La Palma.  We repeatedly image a sample of the 40 most-luminous nearby
(d $<$ 45 Mpc) starburst galaxies in the Ks-band where the extinction
is greatly reduced and the sensitivity of ground-based observations is
still good. These galaxies have been selected such that their far-IR
luminosities are comparable to or greater than those of the two
prototypical nearby starburst galaxies, M 82 and NGC 253.  Galaxies
whose far-IR luminosity is expected to be powered by a population of
old stars or an AGN have been excluded from the sample.  One night of
observations is carried-out every $\sim$3 months. 20-30 galaxy images
per clear night are obtained. More details of the search can be found
in the 'Nuclear SN search' web pages at
http://astro.imperial.ac.uk/nSN.html.
\subsection{Preliminary results from the WHT search}
The total number of Ks images obtained with the WHT so far is
$\sim$110. Archival K-band data having an acceptable seeing and
similar depth to our INGRID images provides a further 10 images.
Thus, our total database currently contains repeat images for 33
starburst galaxies, on average 3.4 epochs per target. However, the
SN detection efficiency (especially within the galaxy nuclear regions)
falls rapidly as seeing quality declines. Therefore, for the analysis
presented here we include only data having seeing better than FWHM =
1.1''. Within this constraint, about 60$\%$ of our INGRID images have
acceptable seeing, reducing the number of targets with repeat images
to 21.  On average, this is $\sim$2.6 epochs of data per target.
\subsubsection{A possible SN in NGC 7714}
Comparison of archival (UKIRT IRCAM3) data with our INGRID images
yielded the discovery of a possible SN with $m_{K}$ = 17.3 only 1 kpc
projected distance from the nucleus of the starburst galaxy, NGC 7714
[18]. From the non-detection of this SN in the H-band we estimated a
lower limit for the extinction towards the SN of Av $\sim$ 6.  Radio
confirmation of this event was also sought.  At radio wavelengths
CCSNe are often still luminous several years from the explosion, with
the most luminous reaching their peak luminosities up to 4 years after
explosion [19].  Therefore, radio follow-up
observations of nuclear SNe may still be feasible long after the SN
has faded below near-IR detectability.  We obtained VLA observations
of NGC 7714 in the L and C-bands on 11 April 2002, but the possible SN
was not detected. The 3$\sigma$ upper limits in the L and C-bands are
0.213 mJy and 0.147 mJy respectively (C. Stockdale, private
communication). These upper limits indicate that the SN was probably
fainter than the type IIn SN 1998S at radio wavelengths.
\subsubsection{Expected number of SN discoveries}
We have made preliminary estimates of the expected number of SN
discoveries from the data collected so far.  To do this we used Monte
Carlo simulations similar to those described in [12]. We assumed that
outside the innermost 100 pc region of our starburst targets all the
SNe brighter than the possible SN in NGC 7714 ($m_{K}$ = 17.3) would
have been detected in the images with seeing better than 1.1". We used
our K-band template light curves for linearly declining "ordinary"
CCSNe and "slowly-declining" CCSNe [12], the amount and distribution of
extinction found for the SNRs of M 82 (Fig.1 right) [12], and the CCSN
rates calculated from the far-IR luminosities of the targets according
to $r_{\rm SN}$ = 2.7 $\times$ 10$^{-12}$ $\times$ $L_{\rm FIR}$/L$_{\odot}$
yr$^{-1}$ [12].  In Fig.3(left) the expected number of SN detections
is plotted as a function of the intrinsic fraction of slowly-declining
events. The results are plotted for a fixed extinction of $A_{V}$ = 10
towards the nuclear SNe, and for an extinction distribution similar to
the SNRs of M~82. For one likely SN detection from the data collected
so far, and an extinction of $A_{V}$ = 10 we obtain an upper limit
($>$90\% confidence) of 30\% for the slow-decliner fraction. However,
assuming an M~82-like extinction distribution yields a most probable
slow-decliner fraction of $\sim$10\%.  In Fig.3(right) the simulation
results are plotted as a function of the extinction towards the
nuclear SNe. Again, assuming one detected event and a slow-decliner
fraction of 30\% indicates a lower limit ($>$90\% confidence) for the
extinction of $A_{V}$ = 10. Assuming a slow-decliner fraction between
5\% (observed outside the nuclear starburst regions) and 30\% yields a
most probable extinction, $A_{V}$, between $\sim$20 and $\sim$30
magnitudes (for one detected SN).
\begin{figure}
\begin{center}
\resizebox{12cm}{!}{
{\includegraphics{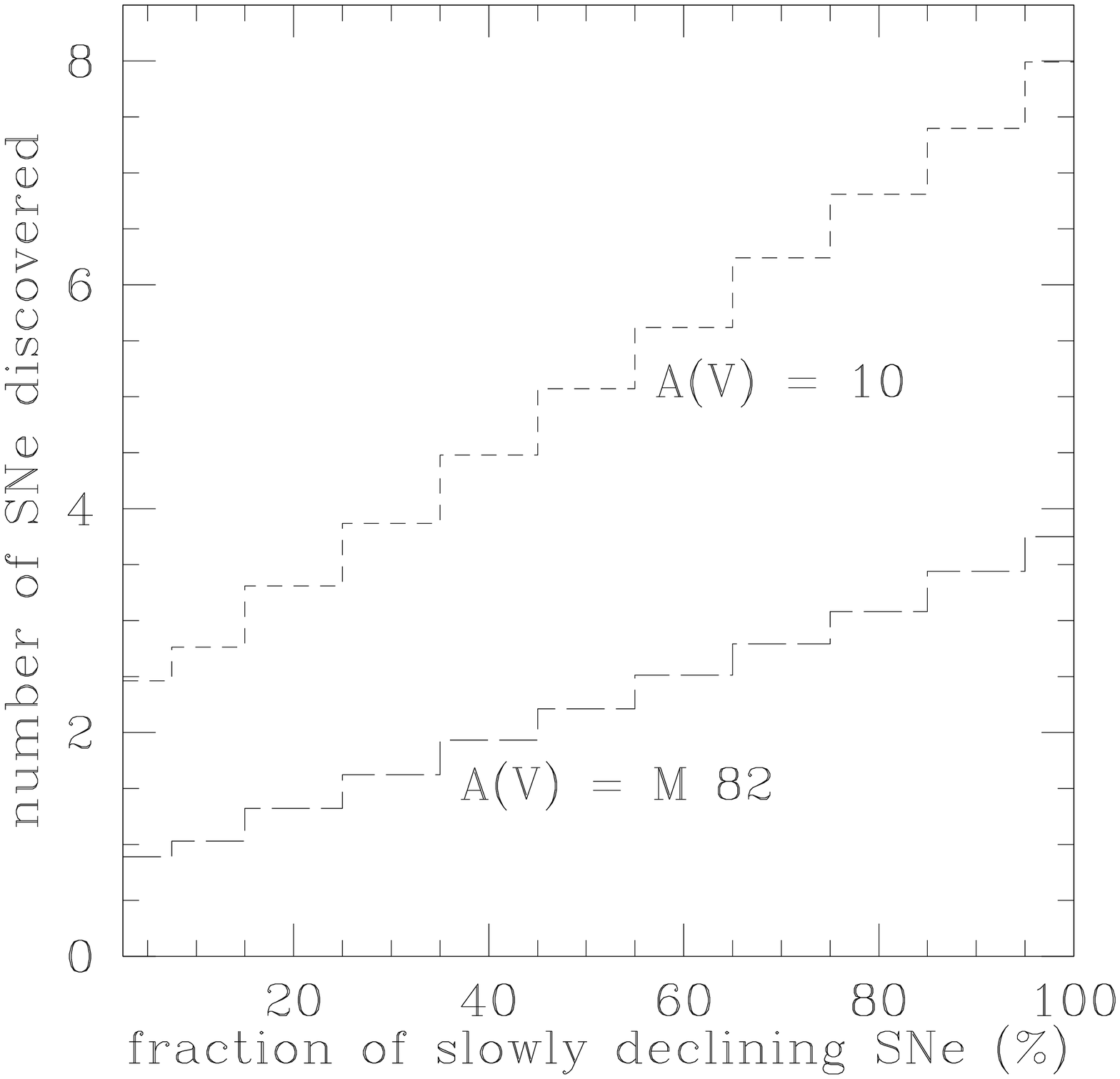}}
{\includegraphics{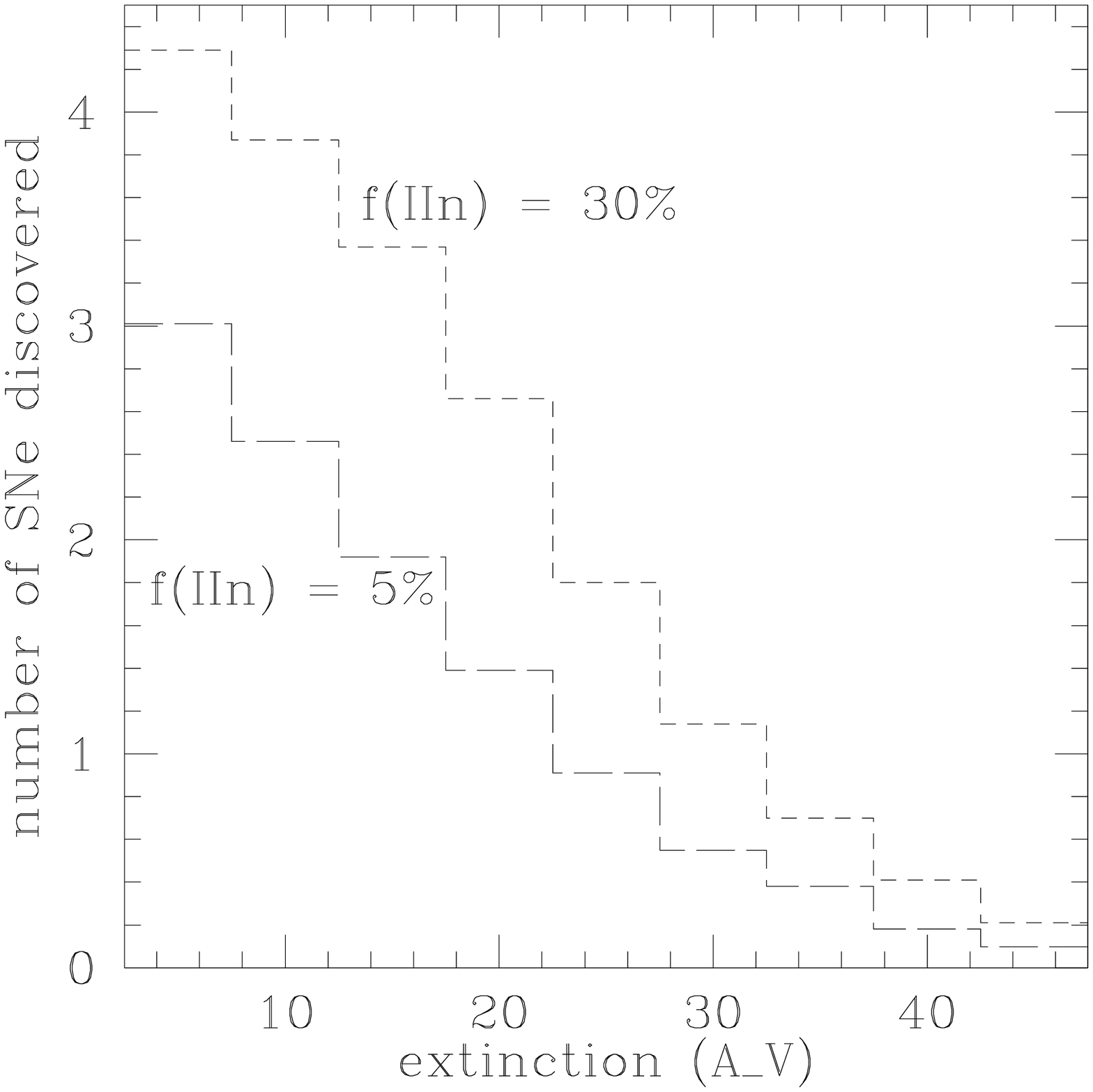}}
}
\caption{The simulated number of SNe discovered with the WHT as a 
function of the slow-decliner fraction (left) and the extinction 
towards the SNe (right). In the left-hand plot, the two different 
histograms correspond to a fixed extinction of $A_{V}$ = 10, and an 
M~82-like extinction distribution towards the SNe. In the right-hand
plot, the histograms correspond to fixed slow-decliner fractions of 5 and
30\%.}
\label{fig1}
\end{center}
\end{figure}
\section{Supernovae in Luminous and Ultraluminous Infrared Galaxies}
Maiolino et al. [2] and Mannucci et al. [20] used an arsenal of
ground-based telescopes (ESO NTT at La Silla, TNG
on La Palma, and AZ61 on Mount Bigelow) to obtain a total of $\sim$230
K-band observations of 46 Luminous Infrared Galaxies (LIRG) (10$^{11}$
L$_{\odot}$ $<$ $L_{\rm IR}$ $<$ 10$^{12}$ L$_{\odot}$) between Oct. 1999
and Oct 2001. Their targets have distances ranging between $\sim$40
and $\sim$300 Mpc, with 28 of them being more distant than
100 Mpc. From these data they discovered two SNe, the type II SN
2001db (d $\sim$ 40 Mpc), and SN 1999gw (d $\sim$ 170 Mpc) with no
spectroscopic typing. These SNe have projected galactocentric (K-band)
distances of $\sim$1.5 kpc (8.5'') and $\sim$3 kpc (3.5'') 
respectively. In addition, during their search two more SNe were
detected in their target galaxies but at optical wavelengths. These
are the type Ia SN 1999gd (d $\sim$ 80 Mpc) and the type IIn SN 2000bg
(d $\sim$ 100 Mpc) with the projected galactocentric distances of
$\sim$7 and $\sim$8 kpc respectively. The extinctions towards the two
optically detected SNe are most likely negligible. Furthermore, the
discovery magnitude of SN 1999gw of $m_{Ks}$ = 17.45 ($M_{Ks}$ =
-18.7) is very close to the peak magnitudes of both 'ordinary' CCSNe [12]
and 'Branch-normal' type Ia SNe [21] indicating a likely small extinction
towards this event (regardless of its type). However, SN 2001db has a
host galaxy extinction of about $A_{V}$ = 5 [2]. In summary, only one
of the four detected SNe has a derived visual extinction of several
magnitudes and is located (projected distance) fairly close to
(but not within) its host galaxy nuclear regions. We also note that this SN
occurred in the nearest (d $\sim$ 40 Mpc) of the 46 target galaxies.
\begin{figure}
\begin{center}
\resizebox{6cm}{!}{
{\includegraphics{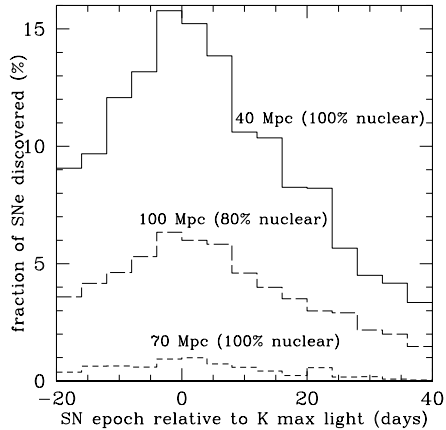}}}
\caption{The fraction of simulated SNe discovered with the NTT
(Mannucci et al. [20]) as a function of the SN age. The different
histograms correspond to three combinations of the host galaxy
distance and the fraction of SNe exploding within the nuclear
regions. A fixed extinction of $A_{V}$ = 30 is assumed towards 
all the events.}
\label{fig1}
\end{center}
\end{figure}

In order to judge the efficiency of the reported Maiolino$/$Mannucci
SN search for host galaxies of different distances, we carried out
Monte Carlo simulations similar to those described in Section
2.1.2. We simulated two scenarios: (1) 100\% of the SNe occur within
the nuclear regions, and (2) 80\% of the SNe occur within the nuclear regions, and 20\%
outside the nuclear regions. We adopted the reported [20] NTT/SOFI
limiting magnitudes for SNe 'on-nucleus' and 'off-nucleus' (outside
the central $\sim$2'', corresponding to $\sim$1~kpc at 100 Mpc distance) of $m_{K}$
= 16.9 and 19.3 respectively. We assumed an intrinsic fraction of
slowly-declining CCSNe of 5\%, and a fixed extinction of $A_{V}$ = 30
towards all the events. In Fig.4 the fraction of simulated SNe discovered is
plotted as a function of the SN epoch. In galaxies at 40 Mpc distance,
$\sim$15\% of the SNe 'on-nucleus' are detectable near maximum light
whereas at 70 Mpc distance the SN detection efficiency falls to less than
1\%, i.e., the survey was not sensitive to highly obscured SNe within
the nuclear regions of target galaxies at $\sim$70 Mpc distance or
further (only 2/46 of the targets were closer than this). However, if we allow 20\% 
of the SNe to occur outside the central regions, about 6\% of the SNe 
near maximum light can be recovered at distances as high as 100 Mpc 
(this includes 18/46 of the targets). Therefore, the average extinction  
of $A_{V}$ $\sim$ 30 derived by Mannucci et al. [20] can only be valid for 
SNe occurring outside the galaxy nuclear regions rather than for all SNe regardless 
of their location within the host galaxy. However, as none of the SNe which were 
detected outside the nuclear regions had such a high extinction, a more plausible reason
for the lack of SN detections would be the concentration of almost all
the SNe within the nuclear (central kpc) regions in the LIRG targets
since this survey was not sensitive to highly obscured SNe in those
regions. Also, the concentration of the luminous radio SNRs (or SNe)
within the innermost nuclear regions in M~82, Arp~220, and other
galaxies support a scenario where $\sim$100\% of the SNe occur within the
nuclear regions in starburst galaxies over a range of IR luminosities.

The large star formation rates (SFRs) within Ultraluminous Infrared
Galaxies (ULIRGs) ($L_{\rm IR}$ $>$ 10$^{12}$ L$_{\odot}$) imply CCSN
rates which are an order of magnitude higher than in nearby
starburst galaxies.  Consequently it may be argued that ULIRGs provide
the best laboratories for the detection and study of CCSNe in the
extreme starburst environment.  The serendipitous VLBI discovery of a
dozen luminous radio SNe/SNRs [13] in the nuclear regions of the
archetypal ULIRG Arp~220 has already demonstrated that nuclear CCSNe
at very late phases can be detected at radio wavelengths. However,
with the unprecedented spatial resolution of the Naos Conica (NACO)
adaptive optics camera on the ESO Very Large Telescope (VLT)
($\times$15 better than previous ground-based studies [2,20], and
$\times$3 better than HST/NICMOS) it is now possible to discover these
SNe shortly after explosion when they are still bright at near-IR
wavelengths. Therefore, we have started a Ks band survey for highly
extinguished CCSNe within the nuclear regions of the nearest ULIRGs,
using the VLT/NACO. We anticipate that the first SN discoveries will
be made during the ESO Period 73, between April 2004 and September
2004.
\section{Discussion}
The WHT SN search data collected by us so far indicates that SNe
within the nuclear (central kpc) regions of nearby (d $<$ 45 Mpc)
starburst galaxies (mostly 10$^{10}$ L$_{\odot}$ $<$ $L_{\rm IR}$ $<$
10$^{11}$ L$_{\odot}$) probably suffer from extinctions higher than
$A_{V}$ = 10, with the most likely average extinction between $A_{V}$
$\sim$20 and 30.  The monitoring of a sample of more distant (mostly d
$>$ 70 Mpc) and more luminous (10$^{11}$ L$_{\odot}$ $<$ $L_{\rm IR}$ $<$
10$^{12}$ L$_{\odot}$) targets recently reported by Mannucci et
al. [20]  indicates that either ${\it(1)}$ $\sim$100\% of the SNe in
LIRGs occur within the nuclear (central kpc) regions, or ${\it(2)}$ if
20\% of the SNe occur outside the nuclear regions then the extinction
{\it towards these off-nuclear SNe} is probably very high, $A_{V}$ $\sim$ 30.  
Only very few obscured SNe (with $A_{V}$ $\sim$ 5) have been detected so
far. However, a number of infrared SN search campaigns have begun in
the past few years.  These include high resolution surveys by Maiolino
et al. using HST/NICMOS and by ourselves using VLT/NACO.  Therefore,
the number of obscured SN discoveries is expected to increase
substantially in the near future. This will {\it eventually} allow
complete SN rates to be estimated for galaxies in the local
Universe. In addition, the discovered SNe will be invaluable as probes
of the extinction in the optically obscured parts of galaxies.

{\bf Acknowledgements}\\ The NOT spectrum of SN 2002hh was observed
and reduced by Jens Andersson, Maiken Gustafsson, P\'all Jakobsson,
Geir \O ye, and Jesper Sollerman. The VLA observations of the possible
SN in NGC 7714 were taken and analysed by Nino Panagia, Dick Sramek,
Christopher Stockdale, Schuyler Van Dyk, and Kurt W. Weiler. The
'Nuclear SN search' (WHT) team includes also Stuart Ryder, Nic Walton,
and Bob Joseph. We also thank the people involved in our 'Supernovae
in ULIRGs' VLT project, in particular Petri V\"ais\"anen and Duncan
Farrah.

% Bibliographic references with the natbib package:
% Parenthetical: \citep{Bai92} produces (Bailyn 1992).
% Textual: \citet{Bai95} produces Bailyn et al. (1995).
% An affix and part of a reference:
%   \citep[e.g.][Ch. 2]{Bar76}
%   produces (e.g. Barnes et al. 1976, Ch. 2).

\end{document}